\documentclass{article}

\usepackage{arxiv}

\usepackage[utf8]{inputenc} 
\usepackage[T1]{fontenc}    
\usepackage{hyperref}       
\usepackage{url}            
\usepackage{booktabs}       
\usepackage{amsfonts}       
\usepackage{nicefrac}       
\usepackage{microtype}      
\usepackage{lipsum}		
\usepackage{graphicx}
\usepackage[square,numbers]{natbib}
\usepackage{doi}

\title{Empirical Formal Methods: Guidelines for Performing Empirical Studies on Formal Methods}

\author{ \href{https://orcid.org/0000-0002-2930-6367}{\includegraphics[scale=0.06]{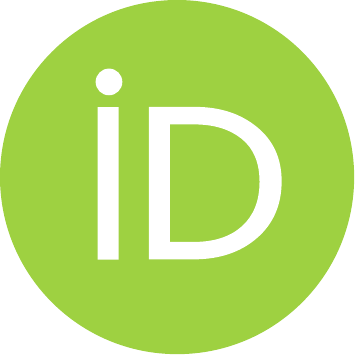}\hspace{1mm}Maurice ter Beek}\\
	Istituto di Scienza e Technologie dell'Informazione ``A. Faedo'' (ISTI)\\
	Consiglio Nazionale delle Ricerche (CNR)\\
	Pisa, Italy, 56126 \\
	\texttt{maurice.terbeek@isti.cnr.it} \\
	\And
	\href{https://orcid.org/0000-0002-0636-5663}{\includegraphics[scale=0.06]{orcid.pdf}\hspace{1mm}Alessio Ferrari}\thanks{Corresponding author: alessio.ferrari@isti.cnr.it} \\
	Istituto di Scienza e Technologie dell'Informazione ``A. Faedo'' (ISTI)\\
	Consiglio Nazionale delle Ricerche (CNR)\\
	Pisa, Italy, 56126 \\
	\texttt{alessio.ferrari@isti.cnr.it} \\
}

\renewcommand{\shorttitle}{Empirical Formal Methods}

\hypersetup{
pdftitle={Empirical Formal Methods: Guidelines for Performing Empirical Studies on Formal Methods},
pdfauthor={Maurice H. ter~Beek, Alessio Ferrari},
pdfkeywords={First keyword, Second keyword, More},
}

\begin{document}
\maketitle

\begin{abstract}
	Empirical studies on formal methods and tools are rare. In this paper, we provide guidelines for such studies. We mention their main ingredients and then define nine different study strategies (laboratory experiments with software and human subjects, usability testing, surveys, qualitative studies, judgement studies, case studies, systematic literature reviews, and systematic mapping studies) and discuss for each of them their crucial characteristics, the difficulties of applying them to formal methods and tools, typical threats to validity, their maturity in formal methods, pointers to external guidelines, and pointers to studies in other fields. We conclude with a number of challenges for \emph{empirical formal methods}.
\end{abstract}


\section{Introduction}


For over two decades, empirical strategies, such as controlled experiments, case studies, surveys, literature reviews, \textit{etc.}, have largely been used to assess software engineering methods and tools, to study software practice, and to summarise research findings~\cite{zhang2018empirical}. However, empirical studies on formal methods (FM), which are mathematically-based techniques and associated tools that typically target the development of demonstrably dependable software, are notably scarce. This has been highlighted already in 2007, by Höfer and Tichy~\cite{hofer2007status}, in their analysis of the status of empirical research in software engineering, and has been further stressed in 2015 by the research agenda of Jeffery \emph{et al}.~\cite{JSAKM15}, calling for a better uptake of empirical methods in FM. A more recent call to arms for the FM community comes from the the manifesto for applicable FM by Gleirscher \emph{et al}.~\cite{GPW21} from 2021. One of the points of the manifesto states that 
\lq\lq [FM] effectiveness should be evidenced. For example, it should be demonstrated (e.g., by means of case studies or controlled experiments) what would have been different if a conventional or non-formal alternative had been used instead.\rq\rq\ Among the expected impacts of the manifesto, it is worth mentioning the \lq\lq Impact on the Conduct, Writing, and Review of Formal Method Research\rq\rq\ and the \lq\lq Impact on the Evaluation of Future Formal Method Research\rq\rq, both calling for case studies, action research, and controlled experiments.
\lq\lq Following these methods would greatly benefit the FM community\rq\rq. Also Huisman \emph{et al}.~\cite{HGM20} recommend to \lq\lq Invest time in industrially-relevant case studies in order to understand what techniques are actually needed for industrially-relevant applications.\rq\rq

Despite these statements, FM research remains focused on developing novel techniques, typically tackling more complex problems or performance issues, and tends to remain a method/tool focused discipline, rather than an evidence-based one. 
Furthermore, by focusing solely on the technical dimension, FM research does not sufficiently take into account human and social factors, which nevertheless affect the usage of FM  tools~\cite{ferrari2020comparing}. In other terms, the discipline of \textit{empirical formal methods} still remains a poorly explored avenue. 
Without demonstrated evidence of effectiveness and applicability, skepticism about FM remains among practitioners, and the industrial uptake of FM is still limited. We argue that, among other factors, this is also hampered by the absence in the FM community of a sufficient knowledge of the available empirical strategies, their fundamental principles, and typical guidelines. 

To address this gap, this paper aims to support future research in \emph{empirical formal methods} with a summary of the main strategies, and a set of guidelines to better apply them in FM. In particular, we consider nine empirical strategies, namely laboratory experiments with software subjects; laboratory experiments with human subjects; usability testing; surveys; qualitative studies---with reference to grounded theory; judgement studies; case studies---arguably including design science and action research; systematic literature reviews; and systematic mapping studies. Though other research strategies exist (cf., e.g., Stol and Fitzgerald~\cite{stol2018abc} for a comprehensive framework), we believe that these can be considered as the most representative and useful for FM researchers. 

In this paper, we first provide an overview of the main ingredients that each empirical study should contain, such as research questions, data collection and analysis procedures, and threats to validity. Then, for each strategy, we summarise: their crucial characteristics, the difficulties of applying them to FM and tools, typical threats to validity, their maturity in FM, pointers to external guidelines, and pointers to studies in other fields. This paper can be used as a concise reference for FM researchers who want to carry out an empirical study, but do not know where to start from---and do not want to incur in typical pitfalls. Our wish is to facilitate the development of an empirical mindset in the FM community.

\section{Fundamental Ingredients}
\label{sec:fundamentalingredients}

Empirical studies are structured research procedures that aim to derive some theory from the observation of phenomena in a study setting. They apply systematic protocols for data collection and analysis, accompanied by validity procedures to reduce researcher bias and mitigate threats to validity. Empirical studies differ for their degree of realism, the ability to generalise outside the study setting, and the ability to isolate the observed phenomena from exogenous factors. However, they all have in common a general structure to keep in mind, which can be useful also for reporting. The fundamental ingredients of this structure are:
\begin{itemize}
    \item \textbf{Research Questions (RQs):} these are statements in interrogative form that drive the research. They are useful as a guideline for the researchers, who has a set of clear objectives to address, but also for the reader. The RQs also typically define the \textit{constructs} of interest, which are the abstract concepts (e.g., formal tools, efficiency, adoption) to be investigated through the research.
    \item \textbf{Data Collection Procedure:} since empirical studies stem from data, these need to be collected, and a systematic and repeatable procedure needs to be established. The data collection procedure specifies which are the data sources, and how data is collected. Data is related to the constructs of interest, as one aims to use the data to measure or evaluate such constructs. 
    \item \textbf{Data Analysis Procedure:} this specifies how the data is elaborated and interpreted to answer the RQs, thus establishing a chain of evidence that goes from data to constructs of interest. In other terms, the data analysis procedure establishes a link between empirical data and RQs. Both data collection and analysis procedures require to consider possible validity threats, and countermeasures to prevent possible threats need to be established and made explicit.
    \item \textbf{Execution and Results:} these specify how data collection and analysis have been carried out, and what is the specific output of these procedures. This part also systematically answers the RQs, based on the available evidence. While in principle data collection and analysis abstract away from concrete data, here the focus is specifically on the data and their interpretation. 
    \item \textbf{Threats to Validity:} these specify what are the possible uncontrolled threats that could have occurred in data collection and analysis, and that could have influenced the observed results. In this part, the researchers should reinstate the mitigation strategies oriented to address typical threats to validity, and acknowledge residual threats. Different threats can typically occur depending on the type of study. Nevertheless, there are three main categories of threats, which in principle apply only to experiments, but that introduce a reasoning framework that can be useful for other types of studies:
    \begin{itemize}
        \item \textit{Construct Validity:} indicates to what extent the abstract constructs are correctly operationalised into variables that can be quantitatively measured, or qualitative evaluated. To ensure construct validity the researcher should show that the constructs of interests are well-defined and well-understood based on existing literature. Furthermore, the researcher should argue about the soundness of the proposed quantitative measures or evaluation strategies. For example, if a researcher wishes to measure effectiveness of a certain tool T, they should present related literature defining the concept of effectiveness, and defining a sound measure for this construct. 
        \item \textit{Internal Validity:} indicates to what extent the researcher has ensured control of confounding external factors that could have impacted the results. These factors includes \textit{researcher bias}, i.e., expectations/inclinations of the researcher that may have impacted the study design (e.g., in a questionnaire definition, or in the data analysis), and any aspect related to subjectivity or context-dependency in the production of the results. Internal validity can also be threatened by time-related aspects, e.g., with a \textit{maturation} effect that could occur when the participants perform multiple tasks one after the other, or with \textit{fatigue} effects due to long experimental treatments. For example, consider the case of comparing two tools A and B on a certain task K. If the subjects first use tool A and then tool B on task K, a learning effect could occur. Indeed, with tool A, they could have learned about task K, and this would have facilitated them in performing the same task when using tool B. To address this issue, the researcher could allocate some subjects only on tool A and others only on tool B. 
        \item \textit{External Validity:} indicates to what extent the results can be applicable to contexts other than the one of the study, or, in other terms, to what extent the results can be considered general, i.e., what is the \textit{scope of validity} of the study. 
    \end{itemize}
\end{itemize}
When selecting a research strategy for an empirical study, and defining a study design, one should consider that there is always a \textit{trade-off} between internal and external validity, and also between the knowledge depth that one could achieve, and the generalisability of the results~\cite{stol2018abc}. For example an experiment should include realistic elements, but its results are typically hardly applicable to real-world cases as the lab context is largely different from a real context---e.g., time constraints, limited realism of models or programs analysed, and the overall \textit{in-vitro}, fictional context. Conversely, a case study is highly realistic, but its results are specific to the organisation in which the case study is carried out, and can hardly be applicable to other contexts. A survey can achieve a high degree of external validity, as a statistically relevant set of subjects are included, but the degree of internal validity is limited, as one can hardly control the subjectivity of the responses. Furthermore, since surveys are oriented to a large number of subjects, the questions should be easy to understand, which limits the knowledge depth that one can achieve, compared, e.g., with case studies or qualitative studies, in which highly informative interviews are carried out, with the possibility of follow-up questions. 

\section{Laboratory Experiments with Software Subjects}
\label{sec:experimentssoftwaresubjects}

\begin{itemize}
    \item \textbf{Definition of the Strategy:} a laboratory experiment is a research strategy carried out in a contrived setting in which the researcher wishes to minimise the influence of confounding factors on the study results. In an experiment with software subjects, the researcher typically compares different tools, algorithms or techniques, to collect evidence, e.g., of their efficiency or effectiveness on a certain representative set of problems.   
    \item \textbf{Crucial characteristics:} in a laboratory experiment with software subjects, one typically defines measurable constructs to be assessed and used for comparison between different software subjects. More specifically, the researcher identifies the constructs of interest, and how these constructs are mapped into variables that can be quantitatively measured, or, if this is not feasible, qualitatively estimated. The constructs of interest are typically strictly connected with the RQs, and the data are the source information that can be used to answer the RQs. Therefore, the researcher also needs to specify how one aims to collect the data associated to the variables. For example, in a quantitative study one may want to focus on the construct of \textit{effectiveness} of a certain tool, with the RQ: \textit{What is the effectiveness of tool T?} If the tool T is designed to find bugs in a certain artefact, this construct can be measured with the variable \textit{bug identification rate} = {number of identified bugs / total bugs}. The data collection strategy could consist in measuring the number of bugs found by tool T on a specific dataset given as input (\textit{number of identified bugs}), which contain a pre-defined set of bugs (\textit{total bugs}). The dataset, also called \textit{benchmark}, should be representative of the set of programs that the tool T aims to verify. If the tool T is designed for a specific type of artefact, then the artefacts should vary across different variables that characterise the artefact: e.g., language, size of the artefact, complexity. It is important to always report the characteristics of the dataset across these salient dimensions. Furthermore, to assess whether the effectiveness of a certain software subject is `good enough', one also needs to define one or more \textit{baselines}, i.e., other tools previously developed, or an artificial predictor (e.g., random, majority class), that can allow the researchers to state that the software subject overcomes the existing baselines for the given dataset. 
     \item \textbf{Weaknesses/Difficulties in FM:} several difficulties may occur when applying this type of strategy in FM. Typically, a software subject is a tool such as, e.g., a model checker or a theorem prover. If the tool requires some interaction with the user, then this can affect the results, as the variable that is measured, e.g., bug identification rate, also depends on the human operator. To address this issue, the experiments should also include design elements that are proper of laboratory experiments with human subjects (cf.\ Sect.~\ref{sec:experimentshumansubjects}). Another pitfall can occur when the tool uses some random or probabilistic principle, and thus the results of the experiment can vary from one execution to the other. In these cases, the tool should be executed multiple times, and confidence intervals, e.g., on its effectiveness or other performance-related constructs, should be estimated and reported with appropriate \textit{p-values}~\cite{box1978statistics}. Furthermore, if one aims to report differences between different tools across multiple runs, appropriate statistical tests, e.g., t-test or Mann–Whitney U test, should be also performed, again reporting p-values and evaluating effect size. Another typical difficulty is identifying a baseline. Indeed, formal tools often target specific fine-grained objectives, e.g., \textit{runtime verification} vs \textit{property proving}, and, in the case of model checkers, can use different modelling languages and different logics for expressing properties. Therefore, the comparison between tools is often hardly possible. In these cases, one can (i)~define simple artificial baselines, against which the tools can be compared; (ii)~restrict the comparison to the subset of the dataset for which a comparison is possible; and (iii)~complement the quantitative evaluation with a qualitative evaluation, involving human subjects in the assessment of the effectiveness of the tool, e.g., with a usability study/judgement study/ (cf.\ Sect.~\ref{sec:usability} and~\ref{sec:judgementstudy}), a questionnaire provided to users after using the tool, or qualitative effect analysis~\cite{kitchenham1997desmet}. 
    \item \textbf{Typical threats to validity:} typical threats are related to the representativeness of the dataset (\textit{external validity}), the soundness of the research design (\textit{internal validity}), and the definition of variables and associated measures (\textit{construct validity}). An inherent threat of this type of study, as for laboratory experiments in general, is the limited realism, as the lab setting is typically contrived and does not account for real-world aspects, e.g., learning curve required to learn a tool, incremental and iterative interaction with tools, or iterative nature of artefact development, which is normally not captured by a fixed dataset. 
    \item \textbf{Maturity in FM:} laboratory experiments with software subjects and in particular tool comparison is relatively mature in FM. Many different competitions exist in which tools are evaluated in terms of performance (evaluation of their usability is rare). Experiments are typically conducted on a representative set of benchmark problems and executed by benchmarking environments like BenchExec~\cite{BeyerLW19}, BenchKit~\cite{KH14}, DataMill~\cite{POZRF16}, or StarExec~\cite{SST14}. The oldest competitions concern Boolean satisfiability (SAT) solvers~\cite{SAT}, initiated three decades ago, and Automated Theorem Provers (ATP)~\cite{CASC}. In 2019, 16 competitions in FM joined TOOLympics~\cite{SWcomp} to celebrate the 25th anniversary of the International Conference on Tools and Algorithms for the Construction and Analysis of Systems (TACAS). For several years now, TACAS and other FM conferences also feature artefact evaluations to improve and reward reproducibility, following their success in software engineering conferences, where they have been introduced over a decade ago~\cite{AEC,AEC2}. A recent survey on FM~\cite[Sect.~5.9]{GBP20} showed that their adoption in industry would benefit a lot from the construction of benchmarks and datasets for FM.
    \item \textbf{Pointers to external guidelines:} guidelines and recommendations for this kind of studies, also called \textit{benchmarking studies}, are provided by Beyer \emph{et al}.~\cite{BeyerLW19} and Vitek \emph{et al}.~\cite{VK12}. Guidelines for combining these studies with other assessment methods are part of the DESMET methodology by Kitchenham \emph{et al}.~\cite{kitchenham1997desmet}. 
    \item \textbf{Pointers to studies outside FM:} an example in the field of automatic program repair is the study by Ye \emph{et al}.~\cite{YeMDM21}, which applies different automatic program repair techniques to the QuixBugs dataset and extensively report also the characteristics of the dataset. In automated GUI testing, Su \emph{et al}.~\cite{su2021benchmarking} compare different tools for the constructs of effectiveness, stability and efficiency. This is also a good example of applying statistical tests to the comparison of different tools. Herbold \emph{et al}.~\cite{herbold2017comparative} also uses statistical tests to compare cross-project defect prediction strategies from the literature on a common dataset. Another representative example, in the field of requirements engineering, is the work by Falessi \emph{et al}.~\cite{falessi2011empirical}. This can be taken as reference in case a set of building blocks need to be combined to produce different variants to be compared. The work is particularly interesting because it also illustrates the empirical principles that underpin the comparison. It should be noticed that the paper does not use a public and representative dataset as benchmark, but a dataset that is specifically from a company, thus including the laboratory experiment in the context of a case study (cf.\ Sect.~\ref{sec:casestudy}). A similar problem of composition of building blocks is considered also by Maalej \emph{et al}.~\cite{MaalejKNS16}, in the field of app review analysis, and by Abualhaija \emph{et al}.~\cite{AbualhaijaASBT20}, in the field of natural language processing applied to requirements engineering. This latter study is particularly interesting as it complements the performance evaluation with a survey with experts.
\end{itemize}

\section{Laboratory Experiments with Human Subjects}
\label{sec:experimentshumansubjects}

\begin{itemize}
    \item \textbf{Definition of the Strategy:} similarly to laboratory experiments with software subjects, an experiment with human subjects is a research strategy carried out in a contrived settings, in which the researcher wishes to minimise the influence of confounding factors on the study results. In these experiments, the typical goal is to evaluate constructs (e.g., understandably, effectiveness) concerning notations, tools, or methodologies, when their evaluation depends on human interaction.
    \item \textbf{Crucial characteristics:} in a laboratory experiment with human subjects, one typically defines measurable constructs to be assessed and used for comparison between different study objects, i.e., notations, tools, or methodologies. More technically, a laboratory experiment with software subjects typically evaluates the effect of certain \textit{independent variables} (e.g., the tool under study, or the experience of the subjects) on other \textit{dependent variables} (e.g., understandability, effectiveness). The researcher thus identifies the constructs of interest, and how these constructs are mapped into variables that can be measured quantitatively or, if this is not feasible, estimated qualitatively. The constructs of interest are typically strictly connected with the RQs, and the data is the source information that can be used to answer the RQs. 
    As for laboratory experiments with software subjects, the researcher also needs to specify how one aims to collect the data associated to the variables. To this end, the researcher typically recruits a set of human subjects, either professional, or, more often, students, and asks them to perform a given task using the objects of the study, i.e., notations, tools, or methodologies. Subjects are typically divided into groups, or \textit{treatments}, the experimental group and the control group. The former uses the object of the study to perform the task. The latter performs the task without using the object of the study. In the task, data is collected concerning the dependent variables, and in relation to the RQs. 
    
    In these experiments, it is typical to refine the RQs into \textit{hypotheses} to be statistically tested, based on evidence collected from the data of the experiment itself. An experiment with statistical hypothesis testing can be seen as two sequential black boxes, an experiment box, and a statistical assessment box. The first box represents the experiment itself which produces data, while the second one represents the actual procedure of hypothesis testing, which uses the data produced by the experiment box to state to what extent one can be confident that, given the data, the effect observed in the data is \textit{not} due to chance. In the experiment box: (i)~the inputs are the so-called \textit{independent variables}, i.e., the variables that the researcher wants to manipulate, for example the type of tool to be used in the treatment; (ii)~the outputs are \textit{dependent variables}, i.e., the variables that represent the constructs that one wishes to observe, e.g., effectiveness, understandability; and (iii)~additional input parameters, e.g., maximum time to execute the task, exercise used in the task. These are the \textit{controlled variables} that are not considered independent variables, but that could influence the effect of independent variables on dependent variables, if not properly controlled, and if their effect is not properly cancelled. In the hypothesis testing box, the inputs are all the data associated to the dependent and independent variables, and  the main outputs are: (i)~the \textit{effect size}, which represents the degree of the observed impact of independent variables on dependent variables; and (ii)~the \textit{statistical significance} of the results obtained, which is given by the \textit{p-value}. The statistical significance roughly indicates how likely it is that the results obtained are due to chance, and not to the treatment. Therefore, lower p-values are preferable, and one normally identifies a significance level, called $\alpha$, often set to $0.05$. When p-value $\leq \alpha$, results are considered significant. In the hypothesis testing box, the output is produced from the input using a certain statistical test (e.g., t-test, ANOVA) that depends on the type of experimental design, and the nature of the variables under study (e.g., rate variables, categorical). 
    \item \textbf{Weaknesses/Difficulties in FM:} applying this type of strategy in FM is made hard by the inherent complexity of most FM. A laboratory experiment typically wants to assess the effectiveness of a tool, but the subjects who will use the tool sometimes also need to be trained on the \textit{theory} that underlies the tool, e.g., formal language, notation, and usage itself. This means that laboratory experiments may need to involve `experts' in FM. However, experts in FM are typically proficient on a specific and well-defined set of approaches (e.g., theorem proving \textit{vs} model checking), and even tools~\cite{steffen2017physics,garavel2019reflections}. Therefore, comparable subjects with similar expertise and in a sufficient number to achieve both statistical power and significance, are hard to recruit, and this makes it difficult to carry out experiments in FM. A possible solution is to focus experiments on fine grained, simple, aspects that can be taught in the time span of a class or a limited tutorial, e.g., a graphical notation or a specific temporal logic. If one wishes to evaluate formal tools, and in particular their user interfaces, it is feasible to perform usability studies (cf.\ Sect.~\ref{sec:usability}). These do not normally require a large sample size (in many settings, 10$\pm$2 subjects are considered sufficient~\cite{hwang2010number,macefield2009specify}, when one adopts specific usability techniques), as they do not aim to assess significance but rather to spot out specific usability pitfalls. If, instead, one wishes to evaluate entire methodologies, it is recommended to decompose them into steps, and design experiments that evaluate one step at the time, e.g., distinguishing between comprehension, modelling phase, verification phase. 
    \item \textbf{Typical threats to validity:} typical threats to validity are associated to \textit{construct validity}, i.e., to what extent the constructs are correctly operationalised into variables, \textit{internal validity}, i.e., to what extent the research design is sound, and all possible factors that could have affected the outcome have been properly controlled, \textit{external validity}, i.e., to what extent the results obtained are applicable to other setting, for example a real-world setting, and \textit{conclusion validity}, which specifies to what extent the statistical tests provide confidence on the conclusion. For conclusion validity, one needs to specify: that the assumption of statistical tests are considered and properly assessed---most of the tests (so called \textit{parametric}) assume a normal distribution of the variables; the value of the statistical power of the tests, which can be estimated based on the number of subjects involved, and that gives an indication of how likely it is that one has incorrectly missed an effect between the variables, while an effect is actually present. Similarly to laboratory experiments with software subjects, an inherent threat is the low degree of realism, as human subjects undertake a task in a constrained environment which somewhat simulates how the task would be carried out in the real world. In other terms, the external validity is inherently limited for these types of study, which tend to maximise internal validity.
    \item \textbf{Maturity in FM:} not surprisingly, given the complexity of many FM, laboratory experiments with human subjects are not extremely mature in FM, but quite some experiments exist. 
    Sobel and Clarkson~\cite{SC02} conducted one of the first quasi-experiments, in an instructional setting, where undergraduate students developed an elevator scheduling system---with and without using FM. \lq\lq The FM group produced better designs and implementations than the control group\rq\rq.
    Debatably~\cite{BT03,SC03}, this contradicts Pfleeger and Hatton~\cite{PH97}, who investigated the effects of using FM in a case study, in an industrial setting, where professionals developed an air-traffic-control information system. 
    They \lq\lq found no compelling quantitative evidence that formal design techniques \emph{alone} produced code of higher quality than informal design techniques\rq\rq, yet \lq\lq conclude that formal design, combined with other techniques, yielded highly reliable code\rq\rq.
    We are also aware of some well-conducted controlled experiments for the comprehensibility of FM like Petri nets~\cite{MMBL93}, Z~\cite{FRF98,SH04}, OBJ~\cite{NW02,CEB05}, and B~\cite{RSPGW07,RSP07}, for a set of state-based (semi-)formal languages like Statecharts and RSML~\cite{ZLL02}, and for domain-specific methods and languages in business process modelling~\cite{SL05,MRC07,RM11}, software product lines~\cite{RFH14,RS14} and security~\cite{LMPT13,LPMR14,LMP17}.
    Further empirical studies on the usability of such FM would be very welcome. The same holds for human comprehensibility and usability of other well-known FM (e.g., Abstract State Machines (ASM), TLA, and calculi like CCS and CSP), even prior to evaluating the effectiveness of tools based on such FM.
    We note that the formalisms of attack trees and attack-defense trees, popularised by Schneier~\cite{Dobb} and formalised by Mauw \emph{et al}.~\cite{AT,ADT}, are claimed to have an easily understandable human-readable notation~\cite{Dobb,securitree}. However, as reported in~\cite{GT17,EHKKPW21}, there have apparently been no empirical studies on their human comprehensibility. Thus, also in this case laboratory experiments with human subjects would be much needed. 
    \item \textbf{Pointers to external guidelines:} Wholin \emph{et al}.~\cite{wohlin2012experimentation} published a book on experimentation in software engineering and the principles expressed in the book also apply to experiments in FM. A practical guide on conducting experiments with tools involving human participants is provided by Ko \emph{et al}.~\cite{ko2015practical}. Guidelines for analysing families of experiments or replications are provided by Santos \emph{et al}.~\cite{santos2019procedure}. To have more insights on experiment design with human subjects, one can also refer to the Research Methods Knowledge Base\footnote{\url{https://conjointly.com/kb/}}~\cite{trochim2022}, an online manual primarily designed for social science, but  appropriate also for experiments in FM. When psychometric is involved because some questionnaire are used to evaluate certain variables, the reader should refer to the guidelines by Graziotin \emph{et al}.~\cite{graziotin2021psychometrics}, specific to software engineering research.  To acquire background on the statistics used in experiments, the handbook by Box \emph{et al}.~\cite{box1978statistics} is a major reference. To focus on hypothesis testing, with clear and intuitive guidelines for the selection of the types of tests to apply, one of the main reference is the book by Motulsky~\cite{motulsky2014intuitive}. The book is, in principle, oriented to biologists, but the provided guidelines are presented in an intuitive and general way, which is appropriate also for an FM readership. It should be noted that, though widely adopted, hypothesis testing has several shortcomings that have been criticised by the research community. Bayesian Data Analysis has been advocated as an alternative option, and guidelines in the field of software engineering have been provided by Furia \emph{et al}.~\cite{furia2019bayesian}.
    \item \textbf{Pointers to studies outside FM:} in software engineering it is quite common to use this strategy, for example to evaluate visual/model-based languages, as done for example by the works of Abrah{\~{a}}o \emph{et al}.~\cite{AbrahaoICG11, abrahao2012assessing}, focused on modelling notations. In the evaluation of methodologies, a representative work is the one by Santos \emph{et al}.~\cite{santos2021family} on test-driven development. When one wants to focus on single specific methodological step, a reference work is the one by Mohanani \emph{et al}.~\cite{mohanani2019requirements}, about different strategies for framing requirements and their impact on creativity. When the focus is human factors, e.g., competence or domain knowledge, a representative work is the one by Aranda \emph{et al}.~\cite{aranda2015effect}, on the effect of domain knowledge on elicitation activities. Finally a comparison between an automated procedure and a manual one for feature location is presented by Perez \emph{et al}.~\cite{perez2020comparing}. 
\end{itemize}

\section{Usability Testing}
\label{sec:usability}

\begin{itemize}
    \item \textbf{Definition of the strategy:} 
    usability testing focuses on observing users working with a product, and performing realistic tasks that are meaningful to them. The objective of the test is to measure usability-related variables (e.g., efficiency, effectiveness, satisfaction), and analyse users' qualitative feedback. It can be seen as a laboratory experiment, but (i)~with a more standardised design; (ii)~with a limited amount of subjects (6 to 12, belonging to 2-3 user profile groups are considered sufficient by Dumas and Redish~\cite{dumas1999practical}); (iii)~collecting both quantitative and qualitative data; and (iv)~whose goal is to identify usability issues, rather than testing hypothesis and achieving statistical significance, which typically require larger samples\footnote{If larger groups of subjects are available, though, quantitative results of usability tests can be evaluated with statistical tests.}.  
    \item \textbf{Crucial characteristics:}
    usability studies can be classified into three main types: (i)~heuristic inspections (or \textit{expert reviews}), in which a usability expert critically analyses a product according to pre-defined usability criteria (cf.\ the list from Nielsen and Molich~\cite{nielsen1990heuristic}), without involving users; (ii)~cognitive walkthrough, in which a researcher goes through the steps of the main tasks that one expects to perform with a product, and reflects on potential user reactions~\cite{mahatody2010state}; and (iii)~usability testing, in which users are directly involved. Here we focus on usability testing, which is also the most common and most studied technique. Usability and usability tests are also the topic of the ISO 9241-11:2018 Part 11 standard~\cite{isousability}. 
    
    In usability tests, the constructs to evaluate, and the related RQs, are pre-defined by the literature, as the researcher typically wants to assess a product according to a set of usability attributes. The usability attributes considered by the ISO standard are \textit{effectiveness} (to what extent users' goals are achieved), \textit{efficiency} (how much resources are used), and \textit{satisfaction} (the user personal judgement with the experience of using the tool). Other possible framing of the usability attributes are the 5E expected from a product, i.e., efficient, effective, engaging (equivalent to satisfaction), \textit{error tolerant}, and \textit{easy to learn} (i.e., time to become proficient with the tool)~\cite{quesenbery2003five}. Holzinger, instead, considers learnability, efficiency, satisfaction, low error rate (analogous to effectiveness), and also \textit{memorability} (to what extent a casual user can return to work with the tool without a full re-training)~\cite{holzinger2005usability}. After a selection of the usability attributes (constructs) that the researcher wants to assess, one needs to define the user profile that will be considered in the test. Test subjects will be selected accordingly and a screening and/or pre-test (i.e., a sort of  demographic questionnaire) will be carried out to assess that the expected profile is actually matched by the subjects. Data collection is performed through the test itself, which is supposed to last about one hour for each subject. A set of task-based scenarios are defined (e.g., installation, loading a model, modifying a model, verification, \textit{etc.}), which the user needs to perform with the tool. Typically, a moderator is present at the test, who will interact with the users, instruct them, incrementally assign tasks and tests, and be available for support, if needed. An observer should also be appointed, who will take notes on user's physical and verbal reactions. Video equipment, as well as microphones, logging computers, logging software (e.g., Inputlog\footnote{\url{https://www.inputlog.net/overview/}}, Userlytics~\footnote{\url{https://www.userlytics.com/}}, ShareX\footnote{\url{https://www.goodfirms.co/software/sharex}}), and eye-tracking devices can be used, depending on the available resources. During the usability test sessions, it is highly recommended to ask the participants to \textit{think aloud}, which means verbalising actions, expectations, decisions, and reactions (e.g., ``now I am pressing the button to verify the model, I expect it to start verification, and to have the results immediately''; ``the tool is stuck, I do not know if it is doing something or not''; ``now I see this result, and I cannot interpret it''). After each task, the user should answer at least the Single Easy Questionnaire (SEQ) test, which means asking how easy was the task in a 7-point scale from 1~--~Very Difficult to 7~--~Very Easy. Other short questions about the perceived time required, and the intention to use can also be asked. After completion of all the tasks, the user typically fills a post-test questionnaire, which measure perception-related variables. Several standard questionnaires exist, e.g.,  SUS (System Usability Scale) and CSUQ (Computer System Usability Questionnaire)---cf.\ Sauro and Lewis for a complete list~\cite{SAURO2016185}. 
    Data from the test are both quantitative and qualitative. Quantitative data include: time on tasks, success or completion rate for the tasks, error rate with recovery from errors, failure rate (no completion or no recovery), assistance requests, search (support from documentation) and other data that can be logged. Results of the post-task and post-test questionnaires, associated to perception-related aspects, are also quantitative. Typical variables evaluated in usability studies, with associated metrics, are reported by Hornbaek~\cite{hornbaek2006current}. Qualitative data include think aloud and observation. Data analysis for quantitative data consists in assessing to what extent the different rates are acceptable, after establishing expected rates beforehand. Since, in the end, what matters is the user perception, the results of post-task and post-test questionnaire are somehow prioritised in terms of relevance, together with qualitative data. For the SUS test, scores go from 0 to 100, and rates above 68 are considered above average. For qualitative data, coding and thematic analysis can be used, similar to qualitative studies (cf. Sect.~\ref{sec:qualitativestudy}). Overall, as for laboratory experiments with human subjects, it is important to establish a clear link between constructs to evaluate and measure variables. Quantitative and qualitative data should be also triangulated to identify further insight. For example, qualitative data may indicate satisfaction in learning the tool, although, e.g., the error rate is high. When more subjects are available for the test, e.g., 30 or more, and one wants to establish statistical relations between variables such as effectiveness, perceived usability and intention to use, one can refer to the Technology Acceptance Model (TAM)~\cite{davis1989perceived}. The relation between usability dimensions and TAM are discussed by Lin~\cite{lin2013exploring}. 
    \item \textbf{Weaknesses/Difficulties in FM:} 
    applying usability tests for FM tools is complicated by the typical need to first learn the underlying theory and principles, and then using an FM tool. It is sometimes difficult to separate difficulty in learning, e.g., a modelling language or a temporal logic, from the usability of a model checker. Therefore, the researcher should first assess the learnability of the theory, e.g., with laboratory experiments with human subjects, and afterwards should evaluate the FM tool. Selected subjects should have learned the theory before using the tool, and usability of FM tools should preferably be evaluated also after some time of usage, so that initial learning barriers have already been overcome by users. If one aims to make a first usability test with users that are not acquainted with FM---which can happen if the target users are industrial practitioners---one can present a tool, perform some tasks and use available post-test questionnaires, like SUS, to get a first measurable feedback, as was done in previous studies~\cite{FMBB21}. When industrial subjects are involved, the difficulty for the researcher is to find problems that are meaningful to the domain of the users, so that these can perceive the potential relevance of the tool. Another difficulty is the typical focus of FM tool developers on the performance of tools, especially for model checkers, with respect to usability aspects, because tools often come from research, which rewards technical aspects instead of user-relevant ones. To be effective, usability tests should be iterative, with versions of an FM tool that are incrementally improved based on the test output. This requires resources specifically dedicated to usability testing. However, if this is not possible, heuristic inspections or cognitive walkthrough should at least be performed on an intermediate version of the tool. Another issuse with FM tools is that these are not websites, but complex systems, which can have several functionalities to test (e.g., simulation or different types of verification as in many model checkers). Given the complexity, one should focus on the most critical aspects to be tested. Another issue with FM tools is the time that is sometimes required to perform verification which makes a realistic test on a complex model often infeasible. In these cases, it is useful to use the so-called Wizard of Oz method (WOZ)~\cite{kelley1983empirical}, in which the output is prepared and produced beforehand, or part of the interaction is simulated remotely by a human.
    \item \textbf{Typical threats to validity:} the typical \textit{construct validity} threats are generally addressed thanks to the usage of well-defined usability attributes and measures. Particular care should be dedicated to the selection of the subjects, so that these are actually representative of the user group considered in the study. Pre-tests or initial screening can mitigate threats. Additional threats to construct validity are related to the way questionnaires are presented. Depending on the formulation of the tests, error of central tendency (the tendency to avoid the selection of extreme values in a scale),  consistent response bias (responding with the same answer to similar questions), and serial position effect (tendency to select the first or final items in a list) need to be prevented. Error of central tendency can be addressed by eliminating central answers, or by asking respondents to explicitly rank items. Consistent response bias can be addressed by using negative versions of the same question, and shuffling the questions. Serial position effect is addressed by shuffling the list of possible answers. To guarantee that the answers to the different questions are a correct proxy of the constructs that one wishes to evaluate, it is also important to perform inter-item correlation analysis~\cite{campbell1959convergent}. \textit{Internal validity} can be hampered by think-aloud activities, which can influence the behaviour of the user, interaction with the moderators, expectations from the tool and possible rewards given after the activity. These threats cannot be entirely mitigated, but the researcher should clarify the following with the user: (i)~what is the status of the tool and the goal of the activity, so that expectations are clear; (ii)~interaction should be minimised; (iii)~it is the tool that is under evaluation and not the user; and (iv)~the reward will be given regardless of the results. Overall, to ensure that the analysis is not biased, it is also important to perform triangulation, that is reasoning about relations between think aloud, observations and post/task and post-test questionnaires. Concerning \textit{external validity}, this can be limited by the low degree of realism given by the test environment, which happens for laboratory experiments. To reduce this, one can perform the test in the real environment, in which the user is typically working, so that interruption, noise and other factors can make the evaluation more realistic. 
    \item \textbf{Maturity in FM:} throughout the years there have been efforts to address usability, but it has by no means become standard practice and many FM tools have never been analysed for what concerns their usability. 
    The PhD thesis of Kadoda~\cite{Kad97} addresses the usability aspects of FM tools. First, using the usability evaluation criteria proposed by Shackel~\cite{Sha86}, two syntax-directed editing tools for writing formal specifications are compared in a practical setting; second, using the cognitive dimensions framework proposed by Green and Petre~\cite{GP96}, the usability of 17 theorem provers is analysed. Hussey \emph{et al}.~\cite{HMC01} demonstrate usability analysis of Object-Z user-interface designs through two small case studies.
    In parallel, there have been many attempts at improving the usability of specific FM through the use of dedicated user-friendly toolsets and the like to hide FM intricacies from non-expert users like practitioners, ranging from the SCR Requirements Reuse (SC(R)$^3$) toolset~\cite{Che98} through the IFADIS toolkit~\cite{LH02,LH06} to FRAMA-C platform~\cite{MPK19} and the ASMETA toolset~\cite{ABGRS19} built around the ASM method. Recently, a preliminary comparative usability study of seven FM (verification) tools involving railway practitioners was conducted~\cite{FMBB21}. 
    The importance of usability studies of FM is confirmed by the recent FM survey by Garavel \emph{et al}., in which over two-thirds of the 130 experts that participated responded that it is a top priority for FM researchers to ``develop more usable software tools''~\cite[Sect.~4.5]{GBP20}.
    \item \textbf{Pointers to external guidelines:} for a prescriptive introduction to usability, the reader should refer to the ISO 9241-11:2018 Part 11 standard~\cite{isousability}. A main reference for usability testing is the handbook by Rubin and Chisnell~\cite{rubin2008handbook}. Nielsen and Molich provide 10 ways to perform heuristic evaluation~\cite{nielsen1990heuristic},  while Mahatody \emph{et al}.~\cite{mahatody2010state} report the state of the art of cognitive walkthrough. Quantitative metrics to measure usability attributes are given by Hornbaek~\cite{hornbaek2006current}. Several resources are made available also via specialised websites, such as \url{https://usabilitygeek.com/}.
    \item \textbf{Pointers to papers outside FM:} A systematic literature review on usability testing, with references scored by their quality, is provided by Sagar and Anju~\cite{sagar2017systematic}. A recent work addressing usability of two modelling tools is presented by Planas~\cite{planas2020uml}. For works using TAM, and focusing on the assessment of attributes related to usability, also including understandability of languages, the reader can consider the works by Abrah{\~{a}}o \emph{et al}.~\cite{abrahao2019assessing,AbrahaoICG11}.
\end{itemize}

\section{Surveys}
\label{sec:surveys}

\begin{itemize}
    \item \textbf{Definition of the Strategy:} A survey is a method to systematically gather qualitative and quantitative data related to certain constructs of interests from a group of individuals that are representative of a population of interest. The constructs are concepts that one wants to evaluate, e.g., usability of a certain tool or developers’ habits. The population of interest (also \textit{target population} or \textit{population}) is the group of individuals that is the focus of the survey, e.g., users of tool T, companies in a certain area, users of tool T from University A vs users from University B, potential users of tool T with a background in computer science, \textit{etc.}
    Surveys are normally oriented to produce statistics, so their output normally takes a quantitative form. Surveys are typically conducted by means of questionnaires, but they can be also carried out through interviews.   
    \item \textbf{Crucial characteristics:} 
    The survey process starts from RQs, and the identification of the constructs of interest just as for the other methods discussed. Then, one needs to characterise the target population, i.e., what are the characteristics of the subjects that will take the survey. Based on these characteristics, the researcher performs \textit{sampling}, which means selecting a subset of subjects that can be considered representative for the population. This is normally carried out with probability sampling, in which subjects are selected according to some probability function (random, or stratified---i.e., based on subgroups of the population) from a sampling frame (i.e., an identifiable list of subjects that in an optimal scenario should cover the entire population of interest, for example the list of e-mail addresses of a company). The sample size required for the survey can be computed considering the size of the target population, desired confidence level, confidence interval and other parameters~\cite{ryan2013sample,rea2014designing}. When designing a survey, one also needs to consider that a relevant portion of the selected subjects, usually about 80-90\%, will not respond to the inquiry. Therefore, to have significant results, one needs to plan for a broad dissemination of the survey, so that, even with a low response rate, the desired sample is reached. If personal data is collected, it is also important to make sure to adhere to the GDPR~\cite{gdpr2016} and to present an informed consent to the subjects. In this phase, it is also important to define the data management plan\footnote{Check \url{https://ec.europa.eu/research/participants/docs/h2020-funding-guide/cross-cutting-issues/open-access-data-management/data-management_en.htm\#A1-template} for a template.}, which includes how the data will be stored and when it will be deleted. After determining the sample size, one needs to design the survey instrument, which can be composed of open-ended and/or close-ended questions. Each type of question has its own advantages and disadvantages, e.g., open-ended questions are richer in information but harder to process, while close-ended questions enable less spontaneous and extensive answers, but are easier to analyse and lead to comparable results between subjects. Regardless of the types of questions selected, a well-designed survey has the following attributes: (i)~\textit{clarity}, i.e., to what extent the questions are sufficiently clear to elicit the desired information;  (ii)~\textit{comprehensiveness}, i.e., to what extent the questions and answers are relevant and cover all the important information required to answer by the RQs; and (iii)~\textit{acceptability}, i.e., to what extent the questions are acceptable in terms of time required to answer them and preservation of privacy and ethical issues. To address these attributes, researchers should perform repeated pilots of the survey instrument, with relevant subjects. If the researcher is not sufficiently confident with the topic of the survey or the type of respondents, it is also useful perform a set of interviews with selected subjects, to better define the questions to be included in the survey instrument. After piloting, the survey can be distributed to the selected sample, and the answers need to be recorded, following the data management plan defined beforehand. 
    Then data is analysed and interpreted. In this phase, researchers should perform some form of \textit{coding} for answers to open-ended questions (cf.\ Sect.~\ref{sec:qualitativestudy}), and should adjust the data considering missing answers. Data analysis and reporting can be performed by first presenting quantitative statistics, with percentages of respondents, possibly followed by more advanced statistical analysis. For example, if RQs concern relationships between variables, statistical hypothesis tests can be performed similar to laboratory experiments with human subjects (cf.\ Sect.~\ref{sec:experimentshumansubjects}). Other advanced methods include Structured Equation Modeling (SEM), which allows researchers to identify relationships between high-level, conceptual and so-called `latent' variables (e.g., background, success, industrial adoption), by analysing multiple observable indicators that can be extracted from the survey (e.g., educational degree and current profession can be considered as indicators of background)~\cite{kaplan2008structural,kline2015principles}. 
    \item \textbf{Weaknesses/Difficulties in FM:} similar to the case of laboratory experiments with human subjects, the main issue is the selection of the participants, i.e., the respondents to the survey. FM experts are an inherently limited population, and each expert is specialised in a limited number of methods or tools. In practice, random sampling is often not practicable, and one needs to recruit as many subjects as possible, thus resorting to the so-called \textit{convenience sampling}. Also, the actual population of FM users, which could be the target of a survey about an FM tool, or about FM adoption in general, cannot be known in advance. Therefore, reasonable assumptions and arguments need to be provided to show that the sample of respondents is actually representative of a certain target population. The FM domain also uses technical jargon, which could make questions and answers not sufficiently clear to a sufficiently wide range of potential respondents. Therefore, in some cases the researchers are constrained to ask only general questions, which however limit the degree of insight that one can achieve. 
    \item \textbf{Typical threats to validity:} the main threats to validity are associated to \textit{construct validity}, which in this case can be actually measured by using different survey questions to measure the same construct, and then performing an inter-item correlation analysis~\cite{campbell1959convergent}. This allows the researcher to discard some items related to a certain construct of interest, because the responses do not appear to be correlated with other items associated to the same construct, or because they are not sufficiently discriminative with respect to other items measuring different constructs. Threats to \textit{internal validity}\footnote{In principle, survey research distinguishes between \textit{validity} (criterion, face, content, and construct) and \textit{reliability}~\cite{taherdoost2016validity}. Here, we use the term \textit{internal validity}, to account for the different validity types, in order to make the explanation more intuitive and consistent with respect to the other strategies described.} concern the way in which the questionnaire is formulated, which could be leading to preferred answers (e.g., all the first answers are checked in a long list of options), and that, if too long, could lead to fatigue effects. The first issue is addressed by shuffling answers between respondents. The second one could be addressed by reducing the length of the questionnaire, and also by shuffling the questions between respondents, so that fatigue effects are compensated. Internal validity can also be affected by systematic response bias, especially in case Likert scales are used. This can occur when similar questions to measure the same construct are always presented in the same affirmative form. The respondent may simply check the same answer, since the questions look similar. To prevent this, the opposing question format is used, in which the same question is asked in positive and negative form. Shuffling the questions also helps at this regard. In survey research, \textit{external validity} should be maximised, as one wants to collect information about an entire population. Claims about the appropriateness of the sample size should be included to support external validity. An additional threat, typical of surveys, concerns \textit{reliability}, which is to what extent similar results in terms of distribution are obtained if the survey is repeated with a different sample on the same population. In practical scenarios, this means that the questions should trigger the same answers if asked to similar respondents, and can be addressed by verifying that the questions are sufficiently clear by piloting the questionnaire with a subset of respondents.  
    \item \textbf{Maturity in FM:} while not particularly mature in FM, some seminal surveys exist. 
    The first systematic survey of the use of FM in the development of industrial applications was conducted by Craigen \emph{et al}.~\cite{CGR95a}. This extensive survey is based on twelve \lq case studies\rq\ from industry and it was widely publicized~\cite{CGR92,CGR95b,Cra95}.
    One of these case studies is also reported in the classical survey on FM by Clarke, Wing \emph{et al}.~\cite{CW96}, together with other \lq case studies\rq\ in specification and verification.
    The comprehensive survey on FM by Woodcock \emph{et al}.~\cite{WLBF09} reviews the application of formal methods in no less than~62 different industrial projects world-wide.
    Basile \emph{et al}.~\cite{BBFGMPTF18} and Ter Beek \emph{et al}.~\cite{BBFFGLM19} conducted a survey with FM industrial practitioners from the railway domain, aimed at identifying the main requirements of FM for the railway industry. 
    Finally, Garavel \emph{et al}.~\cite{GBP20} conducted a survey on the past, present and future of FM in research, industry and education among a selection of internationally renowned FM experts, while Gleirscher and Marmsoler~\cite{GM20} conducted a survey on the academic and industrial use of FM in safety-critical software domains among FM professionals from Europe and North America. 
    \item \textbf{Pointers to external guidelines:} 
    Guidelines and suggestions specific to the software engineering domain are: the introductory technical report by Lin{\aa}ker \emph{et al}.~\cite{linaker2015guidelines}; the comprehensive article, especially covering survey \textit{design}, by Kitchenham and Pfleeger~\cite{kitchenham2008personal}; the article by Wagner \emph{et al}.~\cite{wagner2020challenges}, which has a primary focus on \textit{data analysis} strategies, and related challenges; the checklist by Molleri \emph{et al}.~\cite{molleri2020empirically}; the guidelines specific to \textit{sampling} in software engineering, by Baltes and Ralph~\cite{baltes2022sampling}, especially concerning cases in which probabilistic sampling is hardly applicable. More general textbooks on survey research are: the introductory textbook from Rea and Parker~\cite{rea2014designing}, covering all the relevant topics in an accessible way; the technical book by Heeringa \emph{et al}.~\cite{heeringa2017applied}, specific for data analysis; the extensive book on categorical data analysis by Agresti~\cite{agresti2003categorical}, also covering topics that go beyond survey research in a technical, yet accessible way, and including several examples. For SEM, a primary reference is the book by Kline ``Principles and Practice of Structural Equation Modeling''~\cite{kline2015principles}.
    \item \textbf{Pointers to studies outside FM:} a reference survey, oriented to uncover pain points in requirements engineering, involving several companies across the globe, is the NaPiRE (Naming the Pain in Requirements Engineering) initiative\footnote{\url{http://www.re-survey.org/\#/home}}. The results of this family of surveys have been published by M{\'e}ndez-Fern{\'a}ndez \emph{et al}.~\cite{fernandez2017naming}. Another recent and rigorous survey, using SEM, is the one by Ralph \emph{et al}.~\cite{ralph2020pandemic}, on the effects of COVID-19 on developers' work. A survey using hypothesis testing based on multiple regression models is the one by Chou and Kao~\cite{chow2008survey}, about critical factors on agile software processes. Finally, a survey about modelling practices, also using hypothesis testing but with different types of tests, is the work by Torchiano \emph{et al}.~\cite{torchiano2013}. 
\end{itemize}

\section{Qualitative Studies}
\label{sec:qualitativestudy}

\begin{itemize}
    \item \textbf{Definition of the Strategy:} qualitative studies aim at collecting qualitative data by means of interviews, focus groups, workshops, observations, documentation inspection or other qualitative data collection strategies~\cite{lethbridge2005studying}, and systematically analyse these data. These studies aim at \textit{inducing} theories about constructs, based on the analysis of the data. Constructs and RQs can be defined beforehand, or---less frequently in FM---can emerge from the data themselves. Qualitative studies are typically used when the constructs of interest are abstract, conceptual and hardly measurable (e.g., human factors, social aspects, viewpoints, practices). 
    Qualitative studies include the general framework of Grounded Theory (GT)~\cite{vollstedt2019introduction,corbin1990grounded,glaser1978theoretical,charmaz2006constructing}, which has recently been specialised for the analysis of socio-technical systems~\cite{hoda2022}. 
    \item \textbf{Crucial characteristics:} qualitative studies typically start with general RQs about abstract constructs, and perform \textit{iterations} of data collection and analysis to provide answers to the general RQs. 
    Like surveys, qualitative studies require \textit{sampling} of subjects or objects from which data is collected. While with surveys it is typical to resort to probabilistic sampling, with qualitative studies \textit{purposive sampling} is typically used. With purposive sampling, given the RQ, the researcher samples \textit{strategically}, by selecting the units that, in the given context, are the most appropriate to give different internal perspectives to come to a (locally) complete view and answer the RQ. In qualitative studies, the fundamental characteristic is the qualitative nature of the data analysed, in most of the cases sentences produced by human subjects during interviews.  For example, one could formulate a general RQ: \textit{What are the human factors that characterise the understanding of the notation N?}\footnote{Typically, RQs in qualitative studies, and in particular in GT, are \textit{why} and \textit{how} questions~\cite{hoda2022}, which are oriented to investigate the meaning of the analysed situations. However, \textit{what} questions are also common, especially to induce descriptive theories.}, with associated constructs (i.e., \textit{human factors}, \textit{understanding},  \textit{notation N}). To answer the RQ, one can acquire information through interviews involving novice users of the notation. The interview transcripts will be the qualitative data to be analysed. Data analysis is carried out by means of so-called \textit{thematic analysis}~\cite{braun2006using,cruzes2011recommended}. Thematic analysis aims to identify concepts and relations thereof, based on the interpretation of the data. Thematic analysis makes use of \textit{coding}, which means associating essence-capturing labels (called `codes') to relevant chunks of data (e.g., sentences, paragraphs, lines). The codes represent \textit{concepts}, which are then aggregated into \textit{categories}\footnote{Different terminology is used by different researchers and schools of thought in GT. The terminology used here generally follows from Strauss and Corbin~\cite{corbin1990grounded}. Later, we refer to coding family, which comes from Glaser~\cite{glaser1978theoretical}.}, which in turn can be aggregated and linked to one another. For example, one interviewee may say ``\textit{I often suffer from fatigue when reading large diagrams}''. This can be coded with the labels \textit{fatigue}, \textit{read}, \textit{large diagrams}. Another interviewee may say: ``\textit{I find it hard to memorise all the types of graphical constructs, they are too many, and this makes it frustrating to read the diagrams}''. The labels could be: \textit{construct types}, \textit{read}, \textit{frustration}. The concepts \textit{fatigue} and \textit{frustration} could then be aggregated into the more general category, coded as \textit{feelings}. Once concepts and categories are identified, one can identify relationships (hierarchical, causal, similarity, \textit{etc.}), between concepts and categories. The process is iterative, and through the iterations some concepts and categories may be added or removed. The iterations need to resort to \textit{memos} and \textit{constant comparison}. Memos are notes that the researcher writes to justify codes, reflect on possible relations between concepts/categories, or about the analysis process itself. Constant comparison means comparing the emerging graph of concepts and categories with the data, so that there is clear evidence of the link between the more abstract categorisation and the data. The process of data collection and analysis is normally carried out until \textit{saturation} is reached, i.e., until no further information appears to emerge from the collection and analysis of novel data. The theory, which answers the RQ, is represented by the conceptualisation that emerges from the data. In our example, the theory is a graph of all human factors affecting different dimensions of understanding different aspects of the notation~N. The theory can also be represented by means of different classical coding families, which are typical patterns of concepts and relations thereof~\cite{glaser1978theoretical,vollstedt2019introduction}. This general procedure, which we refer to as \textit{thematic analysis}, is applied as part of the general framework of GT. With GT, data collection and analysis are executed as intertwined and iterative activities, and one applies so-called theoretical sampling to identify subjects to interview or objects to analyse based on the theory that has emerged from the data so far. Here, we do not discuss the GT framework, but we recommend the reader interested in qualitative studies to refer to the guidelines of Hoda~\cite{hoda2022}, which are defined for the software engineering field, and can apply also to FM cases.
     \item \textbf{Weaknesses/Difficulties in FM:}
     An inherent difficulty in applying qualitative research methods in FM is the type of skills and attitude required from a qualitative researcher, which typically takes a constructivist (humans \textit{construct} knowledge through interaction with the environment) rather than a positivist stance (humans \textit{discover} knowledge through logical deduction from observation) in developing their research. This means that while FM practitioners search for proofs, and assume that mathematical objectivity can, and shall be achieved, qualitative research takes subjectivity and contradiction as intrinsic characteristics of reality. For this reason, the type of profile that is fit to do qualitative research in FM, and therefore has FM competence plus a constructivist mindset, is rare. Another difficulty in FM is the limited application of FM in industrial fields. Qualitative studies in FM should be based on interviews and observations of subjects practicing FM in real-world settings. Since these subjects are limited, one needs to resort to observations and interviews in research contexts, where FM are practiced, tools are developed and interaction with industrial partners takes place. These studies should complement the viewpoint of researchers with that of industrial partners, in order to have a complete, possible contrasting view of the subject matter.   
    \item \textbf{Typical threats to validity:} threats to validity in qualitative studies can hardly be categorised according to the classes presented in Sect.~\ref{sec:fundamentalingredients}. Other validity criteria shall be fulfilled, and different categorisations are provided in the literature; cf., e.g., Guba and Lincoln~\cite{guba1994competing} (\textit{trustworthiness} and \textit{authenticity}) vs Charmaz~\cite{charmaz2006constructing} (\textit{credibility}, \textit{originality}, \textit{resonance} and \textit{usefulness}) vs Leung~\cite{leung2015validity} (\textit{validity}, \textit{reliability} and \textit{generalizability}). We refer to Charmaz and Thornberg for a discussion on the topic~\cite{charmaz2021pursuit}. Regardless of the type of classification selected, the researcher should ensure that four main practices are followed, which are oriented to ensure that, despite the inherent subjectivity of qualitative research, interpretations are sound and reasonable: (i)~clearly report the method adopted for data analysis, with at least one complete example that describes how the researcher passed from data to concepts, categories and relations thereof; (ii)~in the results section, report quotes that exemplify concepts/categories and relations thereof; (iii)~perform member checking/respondent validation: the researcher needs to (a)~agree with the participants that what is transcribed and reported is actually what was meant by the participants; and (b)~show the findings to the participants to understand to what extent these are accepted and considered reasonable; and (iv)~perform triangulation: this means looking into multiple data sources (e.g., interviews, observations, documents) to corroborate the findings and involving more than one subject in the data analysis. A reference set of steps for structured triangulation between different analysts is reported in the guidelines by Cruzes~\cite{cruzes2011recommended}.  
    \item \textbf{Maturity in FM:} qualitative studies are not mature at all. We are only aware of~\cite{SH01}, where Snook and Harrison report on five structured interviews, lasting around two hours each, conducted with FM practitioners from different companies, all with some experience of using of FM in real systems (e.g., B, Z, VDM, CCS, CSP, refinement, model checking, and theorem proving). They discuss the impact of FM on the company, its products, and its development processes, as well as their scalability, understandability, and tool support.
    It is worth mentioning that for the aforementioned systematic survey by Craigen \emph{et al}.~\cite{CGR95a,CGR95b}, the authors conducted 23 interviews involving about 50 individuals in both North America and Europe, lasting from half an hour to 11~hours. Moreover, the authors of~\cite{BCKUW00} mention that they interviewed FM practitioners, sponsors, and other technology stakeholders in an informal manner. 
    \item \textbf{Pointers to external guidelines:} guidelines for conducting interviews are provided in the book \textit{Social Research Methods} by Bryman~\cite{bryman2016social}, which also contains a comprehensive introductory manual with a relevant part on qualitative methods, including GT. For observational studies---which belong to the field of  \textit{ethnography}~\cite{sharp2016role}---a relevant reference is Zhang~\cite{zhang2019ethnographic}. A primary reference for coding is the book of  Salda{\~n}a~\cite{saldana2021coding}. For GT, one can refer to the already cited article of Hoda~\cite{hoda2022}---which will be followed by an upcoming manual in the form of a book---and to the guidelines of Stol~\cite{stol2016grounded}. 
    \item \textbf{Pointers to papers outside FM:} 
    Examples of qualitative studies based on interviews are:  the one by {\AA}gren \emph{et al}.~\cite{aagren2019impact}, studying the interplay between requirements and development speed in the context of a multi-case study; Yang \emph{et al}.~\cite{yang2021interview}, on the use of exectution logs in software development; Strandberg \emph{et al}.~\cite{strandberg2019information}, on the information flow in software testing. Example studies using GT in software engineering are: Masood \emph{et al}.~\cite{masood2020real}, about the difference between Scrum by the book and Scrum in practice; Leite \emph{et al}.~\cite{leite2021organization}, on the organisation of software teams in DevOps contexts. 
\end{itemize}

\section{Judgement Studies}
\label{sec:judgementstudy}

\begin{itemize}
    \item \textbf{Definition of the strategy:} a judgement study is a research strategy in which the researcher selects experts on a certain topic and aims to elicit opinions around a set of questions, possibly triggered by some hands-on experience, with the goal of reaching  consensus among the experts.
    \item \textbf{Crucial characteristics:} in judgement studies, RQs cover aspects that require specific expertise to be answered, and for which a survey may not provide sufficient insight, or for which research is not sufficiently mature, like, e.g., \textit{What are the main problems of applying FM in industry?}, \textit{In which way can the use of tool T improve the identification of design issues?}. In judgement studies, the researcher typically selects a sample of subjects that are are considered \textit{experts} on the topic of interest. The results of a judgement study can be used to drive the design of questionnaires to later be elaborated into surveys. For instance, once one has identified the typical problems of FM in industry, these problems can be presented as possible options to a larger set of participants. Data collection is typically qualitative, and it is performed by means of \textit{focus groups}, brainstorming \textit{workshops}, or \textit{Delphi studies}. With \textit{focus groups}, the experts (normally 8 to 10) participate in a synchronous meeting in which they are asked to provide their viewpoint on a topic of interest. Before the meeting, a moderator and a note taker are initially appointed, and recording, possibly with video, is set up. 
    Before the focus group, the experts can be faced with a reflection-triggering task, for example observing a model of a system, playing with a tool interface or a more complex task (e.g., designing a model with tool T). This latter case typically occurs when one wants to evaluate a certain tool involving the opinion of multiple experts, e.g., building on top of the DESMET project methodology~\cite{kitchenham1997desmet}. During the focus group, general warm-up questions are asked, also to elicit the expertise of each expert (e.g., \textit{In which projects did you use FM in industry?}) followed by more specific questions (e.g., \textit{What could be the difficulties of using tool~T in industry?}) and closing with a question for final remarks (e.g., \textit{Do you have something to add?}), after a summary. During focus groups, it is recommended to have a whiteboard, in which the moderator reports the results of the discussion so far. The moderator should make sure that all participants express their opinion, and that consensus is eventually reached---or, if not, contrasting opinions are clearly stated and agreed. Focus groups typically last one hour. If needed, multiple focus groups can be organised in parallel, and participants share the final findings in a plenary meeting. Focus groups can be carried out following the Nominal Group Technique (NGT) designed by Delbecq and Van de Ven~\cite{delbecq1971group}. \textit{Workshops} are meetings that include between 10 and 30 participants, and are similar to focus groups in terms of their goal, i.e., brainstorming opinions and reaching consensus. The main difference is that workshops typically address more general questions, and can include different types of experts, with different degrees of expertise, whereas focus groups consist of more homogeneous participants focused on a more specific topic. Workshops can be carried out through adaptations of the NGT technique~\cite{delbecq1971group}, in which each participant answers a general question using one or more sticky notes (e.g., \textit{What are the problems of applying FM in industry?}). Then the sticky notes are read out loud, explained, attached to a whiteboard by the participants, iteratively grouped and prioritised. In focus groups and workshops, a crucial role is played by the moderator, who needs to ensure that none of the participants overtakes the meeting, and that all participants are able to express their viewpoint. 
    With \textit{Delphi studies}, a large number of experts is normally involved with respect to other methods (i.e., more than 30 subjects) and for longer periods of time (weeks to  months), and the goal is to identify best practices or define procedures, aiming also at quantitatively measuring the consensus. The selected experts are individually asked to express their opinions around a certain problem or question, normally in written form and anonymously. The opinions are then shared with the other participants, and discussion takes place in order to reach consensus, similar to a paper reviewing process. The process typically takes place asynchronously. However, in practice, the discussion can also be carried out by means of a dedicated focus group, depending on the goal and the complexity of the RQs. With Delphi studies, multiple rounds of iterations are carried out to reach consensus. The initial round normally consists of an open question oriented to define the items to be discussed in later rounds (e.g., \textit{What are the best practices for introducing FM in industry?}). The second round can give ratings of relevance or agreement to the different items that have been identified collectively (e.g, \textit{How relevant is it to have an internal contact person having some knowledge of modelling?}). Therefore, in this round, quantitative answers are collected. In a third round, participants can re-evaluate their opinions based on the average results of the group. Therefore, in the end, consensus about, e.g., relevance or agreement, can be measured quantitatively. Focus groups, workshops and initial rounds of Delphi studies typically produce qualitative data. Data analysis in all these cases is carried out with thematic analysis, as described in Sect.~\ref{sec:qualitativestudy}. Quantitative analysis in Delphi studies aims at establishing that about 75\% consensus is reached about the identified items~\cite{keeney2006consulting}.
    \item \textbf{Weaknesses/Difficulties in FM:} no particular difficulties characterise judgement studies in FM, which should therefore actually be encouraged as limited experts are typically available on certain techniques or tools, and the issues under discussion are normally particularly complex, e.g., industrial acceptance or scalability of FM tools.
    One practical difficulty arises with focus groups and workshops in which one needs to record many different voices, and it is not always easy to reconstruct the event from voice recordings only. Furthermore, poor equipment can make it difficult to record all the voices in a room. Video recording can address part of these issues, together with extensive note taking and transcriptions made early after the meeting. 
    \item \textbf{Typical threats to validity:} an inherent threat to validity of judgement studies is the limited generalisability across subjects, given the limited sample, which affects \textit{external validity}. However, an accurate and extensive selection of experts on the topic of interest can improve external validity by allowing generalisability across responses~\cite{stol2018abc}. Other threats are related to \textit{internal validity}, since the results of the study may be biased by dominant, disruptive or reluctant behaviour of participants. These issues can be addressed by the moderator and by ensuring balanced protocols for participation. Concerning \textit{construct validity}, the main issue resides in the communication of questions, and the definition of a shared terminology. Piloting the study, and defining a common vocabulary beforehand can mitigate this issue. As for data analysis, typical threats of qualitative studies apply here. 
    \item \textbf{Maturity in FM:} we are aware of only one judgement study on FM. In~\cite{ferrari2020comparing}, nine different FM tools are analysed by 17 experts with experience in FM applied to railway systems. The study identifies specific strengths and weaknesses of the tools and characterises them by their suitability in specific development contexts.  
    \item \textbf{Pointers to external guidelines:} for focus groups, the book by Grueger and Casey~\cite{krueger2014focus} is a primary reference, while, for a quicker tour on this methodology, one should refer to Breen~\cite{breen2006practical}. A reflection on focus groups for software enigneering is reported by Kontio \emph{et al}.~\cite{kontio2008focus}. For Delphi studies, the initial guidelines have been proposed by Dalkey and Helmer~\cite{dalkey1963experimental}, but over 20 variants exist~\cite{mullen2003delphi}. For the most commonly used guidelines, we recommend to refer to the survey by Varndell \emph{et al}.~\cite{varndell2021use}. For the NGT technique, which can be seen as a hybrid between focus groups and Delphi, the reader can refer to the original article~\cite{delbecq1971group}, to the comparative study between NGT and Delphi by McMillan \emph{et al}.~\cite{mcmillan2016use} or to the simple guidelines by Dunham~\cite{dunham1998nominal}. A good overview of different, group-based, brainstorming techniques is reported by Shestopalov~\cite{shestopalov2019}.
    \item \textbf{Pointers to papers outside FM:} Delphi studies are not common in FM nor in software engineering, whereas they are more frequent in healthcare and social sciences. A good example using the Delphi method is reported by Murphy \emph{et al}.~\cite{murphy2019emergency}, in the field of emergency nursing. An example of usage of the NGT technique is presented by Harvey \emph{et al}.~\cite{harvey2012nominal}. Focus groups are more frequently used in software engineering, typically to involve industrial participants in the validation of prototypical solutions, cf., e.g., Abbas \emph{et al}.~\cite{abbas2022relationship}. An example of a focus group study in software engineering, carried out via online tools, is presented by Martakis and Daneva~\cite{martakis2013handling}. An example of a combination of in-person focus groups and workshops in requirements engineering is presented by De Angelis \emph{et al}.~\cite{de2018requirements}.
\end{itemize}

\section{Case Study, Design Science and Action Research}
\label{sec:casestudy}

\begin{itemize}
    \item \textbf{Definition of the Strategy:} a case study is an empirical inquiry about a certain phenomenon carried out in a real-world context, in which it is difficult to isolate the studied phenomenon from the environment in which it occurs. In the software engineering and FM literature, the term `case study' is frequently misused, as it often refers to retrospective experience reports with lessons learned, or to exemplary applications of a technique on a specific case~\cite{wohlin2021case}. In principle, the researcher does not take an active role in the phenomenon under investigation. When the researcher develops an artefact---tool or method---and applies it to a real-world context, one should design the study as Action Research~\cite{staron2020action} or Design Science~\cite{wieringa2014design}. However, the term \textit{case study} is extremely common, and has established guidelines for reporting~\cite{runeson2012case}. These guidelines are generally applicable also to those cases in which the researcher develops an artefact, applies it to data or people belonging to one or more companies, and possibly refines the artefact based on the feedback acquired through multiple iterations. Therefore, in this paper we will discuss only case study research as unifying framework, including also those cases in which the researcher actively intervenes in the context, as is common in software engineering and FM. 
    \item \textbf{Crucial characteristics:} 
    a crucial characteristic of a case study is the extensive characterisation of the \textit{context} in which the investigation takes place, typically one or more companies or organisations---when more than one are considered, we speak about multiple case study. The researcher needs to clearly specify what is the process typically followed by the company, what are the documents produced, who are the actors involved, which are the tools used to support the process and other salient characteristics. Then, one needs to make explicit the \textit{unit(s) of analysis}, i.e., the case being investigated. The unit can be the entire company, a team, a project, a document, \textit{etc}.\ or sets thereof, in case the researcher wants to perform a comparison between different units. Furthermore, one needs to characterise the \textit{subjects} involved in the research, their profile, as well as the \textit{objects}, e.g., documents or artefacts. As for other research inquiries, a case study starts from the RQs. This is also what mainly differentiates a case study from an experience report, which is typically a retrospective reflection, and it is not guided by explicit RQs. In case studies, RQs typically start from the needs of the company in which the study is carried out. The RQs of a case study can include questions related to the application of a certain artefact, e.g., \textit{What is the applicability of tool`T?}, with sub-questions: \textit{To what extent can we reduce the bugs by using tool~T?}, \textit{To what extent is the performance of tool~T considered acceptable by practitioners?}. In other cases, the RQs can be related to understand the process, e.g., \textit{What is the process of adoption of FM in the company?} or \textit{What is the process of V\&V through FM?} As one can see, RQs in case studies can include both qualitative and quantitative aspects, and therefore quantitative and qualitative approaches are used for data collection and analysis, to answer the RQs. For example, to answer a general RQ such as \textit{What is the applicability of FM in company C?} and associated sub-questions outlined above, one can first measure the performance in terms of bug reduction ensured by the application of FM (\textit{quantitative}) and then interview practitioners to understand if these measures are acceptable (\textit{qualitative}). More specifically, one can first observe the number of bugs in one or more projects carried out without FM, and compare this number with similar projects in which FM are applied---always providing an extensive account of the characteristics of the projects. To understand the perception of practitioners, one can interview them after they have experienced the usage of FM in the projects. Overall, these different types of data contribute to give an answer to the initial RQ. Techniques such as laboratory experiments, usability studies, surveys, qualitative studies and judgement studies can be carried out in the context of case studies. However, one needs to consider the limited data points normally available in case studies, and reasonably adapt the available techniques, applying their principles rather than their full prescriptions, and reduce expectations about generality of the findings. 
    \item \textbf{Weaknesses/Difficulties in FM:} a case study requires active participation from industry, and industrial skepticism is the main issue that FM researchers need to face. Even when industrial partners are willing to be involved, researchers need to spend a considerable time understanding the technical application domain and the characteristics of the projects, so that, e.g., a formal model of a product or component can be designed, and the expected properties can be correctly stated. In principle, the researcher should not take part in the application of FM, but this is often impossible, given the complexity and limited maturity of the interface of many a tool. Therefore, it is common to have an interaction scheme in which the researcher develops and verifies formal models, with the support of a main technical contact point from the company, who ensures that the models are faithful to the real system. This communication loop however, should always be completed with a larger involvement of practitioners, who need to assess and confirm a larger applicability of FM in the company, and in general provide some form of structured feedback, e.g., via surveys or interviews. In practice, it is also hard to measure some possibly relevant variables, such as the learning curve of FM and actual bug reduction, throughout projects that can last for years and have several dimensions of complexity (process phases, change of people involved, budget, time constraint, \emph{etc}.). To address this, it is acceptable to use data (e.g, requirements) from previous projects as a main source, and to consider substantial portions of products, instead of entire ones. The researcher somehow sacrifices the realism of the case study in favour of something more manageable, keeping in mind that one should always involve practitioners in a structured reflection about possible consequences of the observed case in a real-world environment. Finally, one needs to consider that case studies deal with proprietary data, which companies are typically not inclined to disclose. This problem can be addressed by providing convincing examples and portions of data, and by setting the terms of the collaboration with the company beforehand, including intellectual property management that clearly states what can be published. 
    \item \textbf{Typical threats to validity:} case study research is typically evaluated according to \textit{construct}, \textit{internal} and \textit{external} validity. Other types of validity, e.g., from qualitative studies or laboratory experiments are considered, when these strategies are used in the context of case studies. \textit{Construct validity} is concerned with the general constructs and associated measurement instruments used in the study. The researcher should comment on their soundness. \textit{Internal validity} is typically concerned with the list of contextual factors that could have impact on the results. As the characteristics of the subjects and objects involved are specific to the case, and since context and phenomena are hard to be separated in a case study, one needs to provide reasonable mitigation measures to reach some form of objectivity. For example, considering more than one single product or component, and involving multiple practitioners, also with multiple roles and competences. \textit{External validity} is also inherently limited in case studies. However, one should report and discuss what are the dimensions of the case that could make its results generalisable to other contexts, according to the principles of case-based generalisation~\cite{wieringa2015six}. For example, safety-critical companies follow highly structured and comparable processes, and railway and avionics are basically oligopolies, following very similar practices. Therefore, a case study in one domain could inform a company in a similar domain, having a comparable degree of maturity. The possibility to generalise, especially based on similarity, drives the need to provide an extensive account of the company or organisational context in a case study paper.  
    \item \textbf{Maturity in FM:} not surprisingly, given the difficulties mentioned earlier, case studies and the like are not very mature in FM. However, next to the aforementioned case study by Pfleeger and Hatton~\cite{PH97}, who investigated the effects of using FM in an industrial setting in which professionals developed an air-traffic-control information system, there are some examples of case studies developed by academics in close collaboration with practitioners---and also partially carried out inside the companies they work for~\cite{GCM98,CCCDHMMMM18}. 
    In particular the railway domain contains a fair number of case studies on applying FM~\cite{LFFP11,ferrari13metro,BQJG15,HP16,CLMPM19}, among which one of the best known success stories of applying FM in industry~\cite{BBFM99}.
    \item \textbf{Pointers to external guidelines:} the primary reference for case study research is the book by Runeson~\cite{runeson2012case}, including also several examples. The reference for action research is the recent book by Staron~\cite{staron2020action}. We recommend referring to action research when the goal is the actual transformation in the company---e.g., through the introduction of an FM tool---and both researchers and practitioners are active in this transformation, e.g, the former as instructors and the latter as trainees. Finally, for design science, one can refer to the books of Wieringa~\cite{wieringa2014design} and Johannesson and Perjons~\cite{johannesson2014introduction}. We recommend referring to design science guidelines when the focus is on the FM tool or interface to be developed. 
    \item \textbf{Pointers to papers outside FM:} an example of a case study in the strict sense, i.e., without intervention of the researcher, is the one by Britto \emph{et al}.~\cite{britto2020evaluating}. The study focuses on on-boarding of software developers and, as is common, here the unit of analysis is a single company. A multiple case study, concerning the usage of DevOps in five companies, is presented by Lwakatare \emph{et al}.~\cite{lwakatare2019devops}.
    Another reference case is reported by Tomasdottir \emph{et al}.~\cite{tomasdottir2018adoption}. The study is about the usage of a static analysis tool by open-source developers. The case is interesting as the unit of analysis is a tool and not a company, and multiple data sources are used, as required by case study guidelines. An iterative case study, in which the researcher performs limited intervention, is presented by Ferrari \emph{et al}.~\cite{ferrari2018detecting}. The study concerns the incremental development of a natural language processing tool for defect detection in requirements documents. An example paper following action research guidelines is \cite{ochodek2020recognizing}, about the application of a machine-learning technique to detect violation of coding guidelines.  Finally, for a reference using the design science paradigm, the reader can refer to Manzano \emph{et al}.~\cite{manzano2021method}. 
\end{itemize}

\section{Systematic Literature Reviews and Mapping Studies}
\label{sec:systematicstudy}

\begin{itemize}
    \item \textbf{Definition of the strategy:} Systematic Literature Reviews (SLRs) and Systematic Mapping Studies (SMSs) are secondary studies (i.e., analysing other empirical research papers, i.e., primary studies) systematically conducted using search engines of digital libraries. These studies are oriented to ensure \textit{completeness}, in terms of surveyed literature, and \textit{replicability}, thus following well-defined guidelines, and carefully reporting the study protocols. SLRs are typically oriented to survey and summarise \textit{findings} from previous research (e.g., identify reported industrial problems with model-checking tools). SMSs mainly aim to scope a research field, identifying relevant dimensions and categorising the literature accordingly (e.g., summarise the literature about theorem proving for safety-critical systems). However, these studies follow similar guidelines and protocols, and SLRs typically include also a categorisation of existing studies, as do SMSs. 
    \item \textbf{Crucial characteristics:} the main RQ of an SMS typically aims to identify the relevant dimensions of a certain field, e.g. \textit{What are the relevant dimensions characterising the literature about FM in avionics?} The RQ is then decomposed into RQs about the demographic distribution of the studies (years, venues, \emph{etc}.), the type of studies (case study, surveys, lab experiments, \emph{etc}.), and other aspects, e.g, \textit{What techniques are used?} and \textit{What tools are used?} These questions are typically answered quantitatively, providing statistics about the frequency distribution of the studies. Besides these RQs, which are shared with SMSs, SLRs also provide more in-depth analyses, answering RQs related to the effect of a technology (\textit{What is the effect on the avionic process of introducing model checking?}), its cost (\textit{What is the cost of introducing model checking in avionics?}), its performance (\textit{What is the performance of model-checking tools in avionics?}) and other aspects that are considered relevant. Answering these questions can require qualitative analyses, conducted using thematic analysis and coding (cf.\ Sect.~\ref{sec:qualitativestudy}). An SMS/SLR needs to be justified, i.e., one should provide an account of related reviews that do not address the same RQs, or do not address them systematically (e.g., systematic studies exist on FM for railways, but not for avionics; studies exist on FM for avionics, but are not systematic, or are outdated). Therefore, typical constructs of SMSs/SLRs are the salient characteristics that the researcher wants to extract from the existing literature (e.g., demographic information, type of study, performance, \emph{etc}.). 
    SLRs/SMSs start from a search string to be used as input for the search engines of specialised digital libraries, which for computer science are IEEE eXplore, Scopus, Web of Science,  SpringerLink and ACM Digital Library. The search string is composed of terms that are relevant to the main RQ, connected with AND/OR logical operators. The search conducted via the search engines is called \textit{primary search}. The search string should be able to identify as many relevant studies as possible, and should limit the irrelevant studies retrieved. To this end, pilot searches need to be performed, possibly adding words to the string, if the researcher identifies terminological variations that are typically used in the literature and enable a more focused search. After retrieving the studies, the researcher needs to select them according to a set of inclusion/exclusion criteria (e.g., removing short papers, removing other secondary studies, removing low quality studies), also removing duplicates that can emerge since multiple search engines are used. The selection activity is typically based on a screening performed on titles and abstracts. To improve replicability, this activity is conducted in parallel by multiple researchers, and disagreement in the selection is resolved through dedicated meetings. Throughout the activities, it is important to always keep track of the number of studies retrieved and selected, and all the search strings used, as different variations of the string may be needed depending on the search engine. 
    To facilitate the removal of duplicates, and the tracing of the whole process, specialised tools like, e.g., Zotero\footnote{\url{https://www.zotero.org/}} or Mendeley\footnote{\url{https://www.mendeley.com/}}, can be used for support. After the selection, researchers can perform a quality assessment of the studies, according to a dedicated rubric to be defined beforehand. This can be carried out to further limit the number of included studies, to report about their  quality, or to answer a subset of the questions considering only high-quality studies. Once the studies have been selected, one performs a \textit{secondary search} to ensure completeness. This means performing forward/backward snowballing (i.e., looking for cited papers, and citing papers, e.g., through Google Scholar), analysing the literature of prominent authors or screening the papers from relevant venues. After the paper selection process has been completed, the researchers extract information according to extraction schemes. These are defined based on the RQs and, in some cases, already outline the predefined options (e.g., the extraction scheme for `types of studies' would have `case study', `survey', \textit{etc}.\ as options). When the options are not clear, one initially extracts relevant text from the paper and then identifies the options after thematic analysis. As for the study selection task, this activity should be performed in parallel by multiple subjects. The results of SMSs/SLRs are reported in the form of plots (for frequency-related RQs) and tables (for qualitative RQs). An SMS/SLR is not a mere summary of the literature. It crucial to provide a contribution based on focuses and gaps in the literature, thereby illustrating further avenues for research and providing recommendations to the community.  
    \item \textbf{Weaknesses/Difficulties in FM:} SMSs/SLRs in FM share the typical difficulties of other fields like, e.g., software engineering~\cite{brereton2007lessons,kitchenham2009systematic}. These include the limitations of search engines of digital libraries, and the difficulty of assessing relevance from the title and abstract alone. Additional complexity dimensions are related to the \textit{publication bias}, i.e., certain evidence is not published in scientific venues. To address this issue, multivocal literature reviews with systematic retrieval of \textit{grey literature}, which includes non-peer reviewed articles, such as blog posts, websites, news articles, white papers and similar is recommended~\cite{garousi2016need,garousi2019guidelines}. Another difficulty, also raised in different points of this paper, is the difficulty of having expertise in multiple FM, which limits the ability to fully understand papers that are within the scope of the RQs, but that cover topics that the researchers do not know in detail. This can be addressed with the involvement of researchers with multiple backgrounds. The reporting of empirical studies in FM does not follow a consistent/standardised structure, and this make it difficult to compare and assess them. The incorrect naming of research strategies, typically `case study' to refer to examples, is also complicating the evaluation, as different quality criteria and extraction schemes are used depending on the study type.
    \item \textbf{Typical threats to validity:} threats to validity in SMSs/SLRs are partitioned into \textit{study selection validity}, \textit{data validity} and \textit{research validity}. Study selection validity concerns threats in search and filtering, and includes bias in search string construction, the selection of digital libraries and primary studies. Typical mitigations are strategies for cross-checking and study selection involving multiple researches, best effort to achieve completeness through secondary searches and appropriate justification for the selection of digital libraries. Data validity concerns threats in data extraction and analysis, and include publication bias (i.e., evidence may exist but it is not published in scientific venues), bias in data extraction schemes and subjectivity in classification and extraction. Mitigation strategies are the additional search for grey literature, the reuse of schemes adopted by previous studies, and again, cross-checking involving multiple researchers. Research validity is concerned with threats to the design as a whole, including coverage of the RQs, replicability and generalisability. The threats can be mitigated by sharing the protocol, clarifying the gap with previous reviews and performing a broad search (e.g., without a starting date for the search time interval). 
    \item \textbf{Maturity in FM:} maturity of SMSs and SLRs is increasing, as witnessed by two SMSs and two SLRs published this year.
    As far as we know,~\cite{FB22} is the first SMS on FM focusing on applications of FM in the railway domain. It considers 328 high-quality studies published during the last 30~years, which are classified according to their empirical maturity, the types of FM applied and railway specific aspects, identifying recent trends and the characteristics of the studies involving practitioners. 
    Moreover,~\cite{FOC22} is the first SLR on the use of FM to evaluate security and energy in IoT, including wireless sensor networks, which can be used as a guide for which FM and tools to use. It considers 38 high-quality studies from a period of 15~years and the findings include a clear predominance of the use of manual proof methods, Dolev-Yao-like attack models, and the AVISPA tool~\cite{ABBCCCDHKMMORSTVV05}.
    Furthermore,~\cite{MM22} is claimed to be the first SLR on FM focusing on the security requirement specification. It considers 88 studies from the last 20~years and it is observed that model checking is preferred over theorem proving, while it remains a research challenge to effectively use FM in a cost-effective and time-saving way.
    Finally, as far as we know,~\cite{ZTKS22} is the first SMS on the use of FM, including semi-formal methods, during the requirements engineering of industrial cyber-physical systems. It considers 93~studies from the last decade and it is shown that safety and timing requirements have been extensively analysed and verified, but there is a lack of work on key phases like requirements elicitation and management, while also the adoption of industrial standards is largely missing and so are methods to handle the currently critical concerns of privacy and trust requirements. 
    Also worth mentioning are some earlier studies, in particular in specific application domains~\cite{DD12,WIIA12,BGM18,MKE18,RFNV21}, as well as on teaching FM~\cite{Zhu19}.
    \item \textbf{Pointers to external guidelines:} Kitchenham~\cite{kitchenham2004procedures} provides the primary reference for conducting SLRs in software engineering and the guidelines are generally appropriate also for SMSs. For SMSs, however, the guidelines from Petersen~\cite{petersen2008systematic,petersen2015guidelines} are a valid alternative. For guidelines on how to select and assess an `optimal' search string, the interested reader can refer to Zhang~\cite{zhang2011identifying}. For guidelines on how to perform snowballing, the reader should refer to Wholin~\cite{wohlin2014guidelines}. For multivocal literature reviews to include grey literature, the reader should refer to Garousi \emph{et al}.~\cite{garousi2019guidelines}. Guidelines for reporting threats to validity are made available by Ampatzoglou \emph{et al}.~\cite{ampatzoglou2019identifying}. Finally, guidelines to update SLRs are discussed by Wohlin \emph{et al}.~\cite{wohlin2020guidelines}.
    \item \textbf{Pointers to papers outside FM:} a recent example of an SLR is the work by D\k{a}browski \emph{et al}.~\cite{dkabrowski2022analysing}. Another example, which also includes a quality checklist, is the one by Bano and Zowghi~\cite{bano2015systematic}, about the practice of user involvement in software development. A good SMS, on software engineering for AI, is the one by Martinez \emph{et al}.~\cite{martinez2022software}. Another example, with a rigorous evaluation of agreement between researchers, is the work by Horkoff \emph{et al}.~\cite{horkoff2019goal}, about goal-oriented requirements engineering. Finally, for multivocal literature reviews, a representative example is the study by Garousi \emph{et al}.~\cite{garousi2017software}. A more recent work, using reference guidelines~\cite{garousi2019guidelines}, is the one by Sheuner \emph{et al}.~\cite{scheuner2020function}.
\end{itemize}



%



\section{Discussion and Conclusion}
Research in FM has traditionally focused on developing novel techniques, improving the performance of FM tools, and tackling ever complex problems. On the other hand, evidence-based, empirically grounded studies are currently limited. To foster a better uptake of \emph{empirical formal methods} in the community, this paper presents a summary of the main empirical research strategies as they are defined in software engineering,  and specifically adapted for their application in FM. Of course, several challenges still exist. In particular: the complexity of the theory behind FM tools, which leads to difficulties in performing usability testing or experiments with human subjects with a sufficient background; the wide differences among tools, which makes it hard to compare them on real-world benchmarks; the development status of many tools, which are not sufficiently mature to be evaluated by industrial practitioners, thus hampering case study research; the fact that FM experts are often specialised on a single tool, thereby making it hard to find sufficient subjects to perform an in-depth survey about a certain tool, and also an experiment with human subjects having comparable backgrounds. Some recommendations on how to overcome these issues in practice are presented throughout the paper. Our study represents a reference guide for researchers who want to approach a FM problem with an empirical mindset, and want to soundly evaluate FM, their tools, or systematically study FM practice and literature. 

\bibliographystyle{unsrt}
\bibliography{references}  






\end{document}